\documentclass[acmtog]{acmart}
\usepackage{color, colortbl}
\usepackage{booktabs}
\usepackage[TABBOTCAP]{subfigure}
\usepackage{makecell}
\usepackage[ruled]{algorithm2e} 
\usepackage{parskip}
\usepackage{amsmath, amsthm,bbm}
\usepackage{txfonts}
\usepackage{hyphenat}
\usepackage{multirow}
\usepackage{subfloat}
\usepackage{hhline}
\usepackage{wrapfig}
\acmPrice{15.00}

\setcitestyle{square}

\setcopyright{none}
\renewcommand\footnotetextcopyrightpermission[1]{} 
\begin{document}

\title{NeuralDrop: DNN-based Simulation of Small-Scale Liquid Flows on Solids}

\author{Rajaditya Mukherjee}
\affiliation{%
  \department{Department of Computer Science and Engineering}
  \institution{The Ohio State University}}
\email{mukherjee.62@osu.edu}

\author{Qingyang Li}
\affiliation{%
  \department{Department of Computer Science and Engineering}
  \institution{The Ohio State University}}
\email{li.7580@osu.edu}

\author{Zhili Chen}
\affiliation{%
  \institution{Adobe Research}}
\email{ zlchen@adobe.com}

\author{Shicheng Chu}
\affiliation{%
  \department{Department of Computer Science and Engineering}
  \institution{The Ohio State University}}
\email{chu.386@osu.edu}

\author{Huamin Wang}
\affiliation{%
  \department{Department of Computer Science and Engineering}
  \institution{The Ohio State University}}
\email{whmin@cse.ohio-state.edu}

\renewcommand{\shortauthors}{Wang}
\newcommand{\comm}[1]{{ \textcolor{red}{#1} }}
\newcommand{\raj}[1]{{ \textcolor{green}{#1} }}

\begin{teaserfigure}
\centering
\subfigure[Before simulatoin]{\includegraphics[width=3.5in]{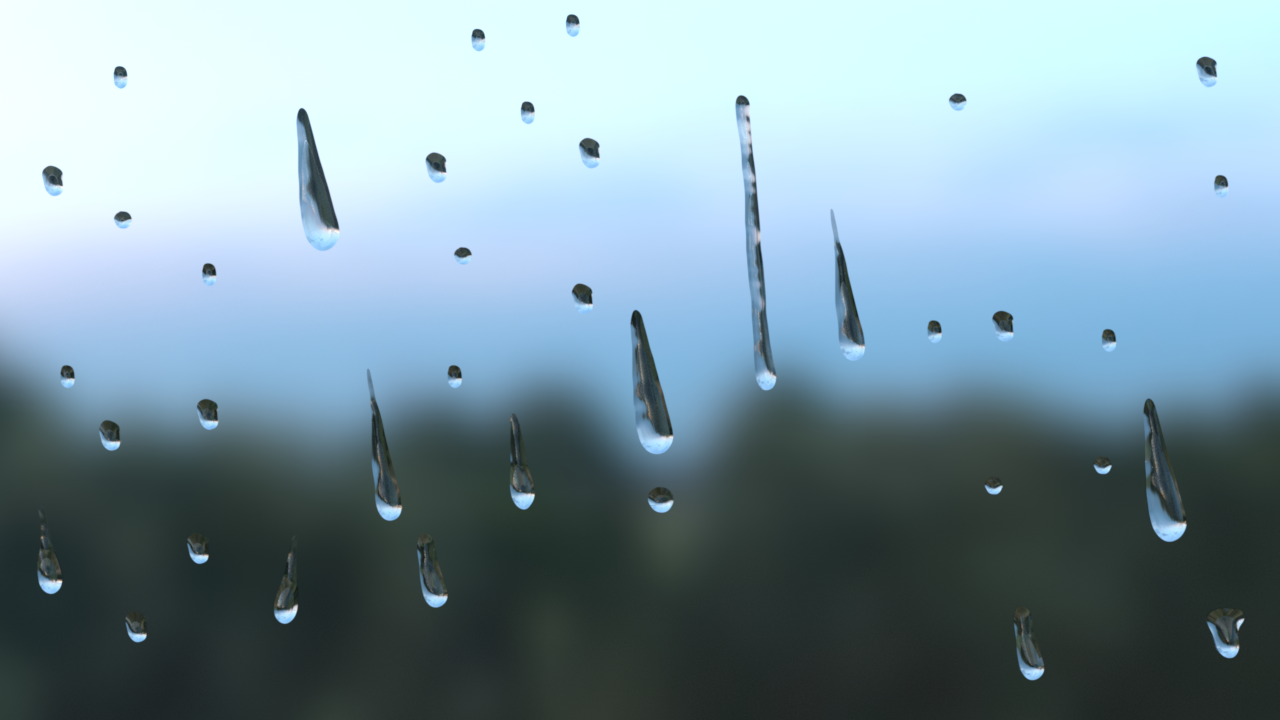}}
\subfigure[After simulation]{\includegraphics[width=3.5in]{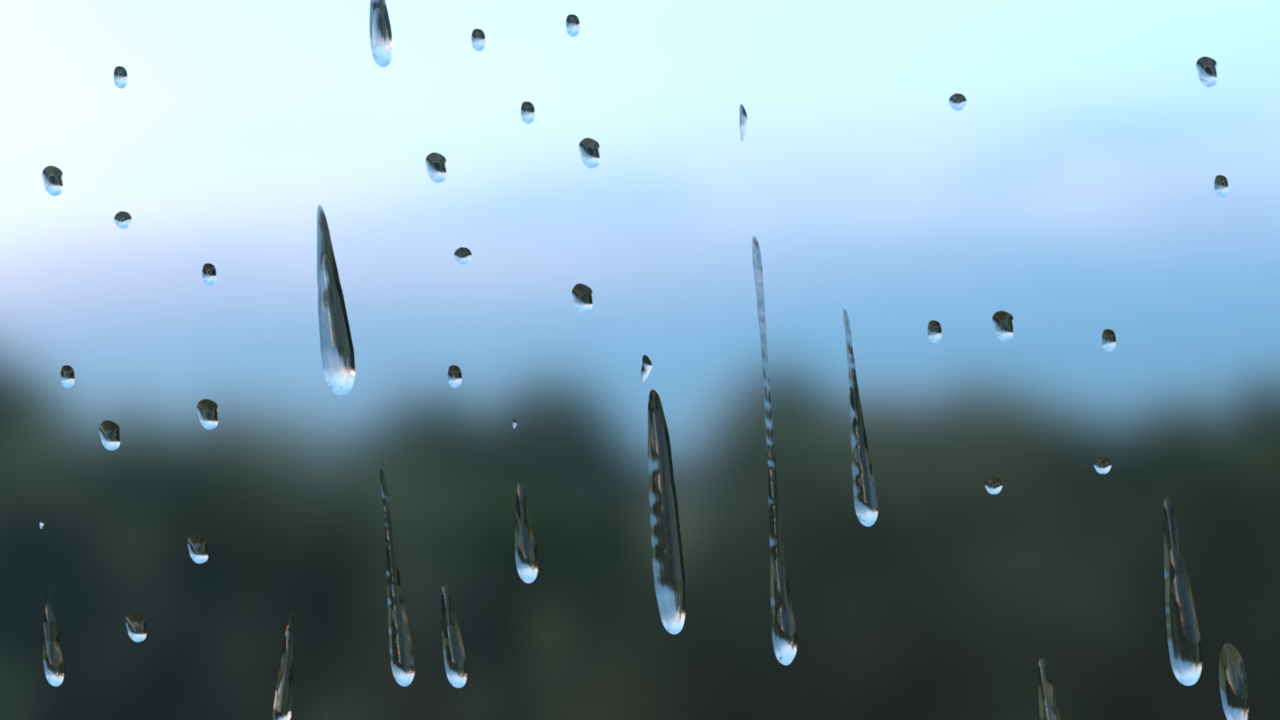}}
\vspace{-0.1in}
\caption{A window example.  The scene in this example contains 50 liquid drops simulated by our novel learning-based liquid simulator, specifically designed for small-scale flows on solid surfaces. The key component of this simulator is LSTM-based recurrent neural networks, trained by real-world flow data.  Using the neural networks, our simulator efficiently predicts the motion of every liquid drop at the next time step in 0.1s.   Integrated with geometric operations, the simulator realistically simulate complex liquid flow effects under surface tension, as shown in (b). }
\vspace{0.16in}
\end{teaserfigure}

\begin{abstract}
Small-scale liquid flows on solid surfaces provide convincing details in liquid animation, but they are difficult to be simulated with efficiency and fidelity, mostly due to the complex nature of the surface tension at the contact front where liquid, air, and solid meet.  In this paper, we propose to simulate the dynamics of  new liquid drops from captured real-world liquid flow data, using deep neural networks.  To achieve this goal, we develop a data capture system that acquires liquid flow patterns from hundreds of real-world water drops.  We then convert raw data into compact data for training neural networks, in which liquid drops are represented by their contact fronts in a Lagrangian form.  Using the LSTM units based on recurrent neural networks, our neural networks serve three purposes in our simulator: predicting the contour of a contact front, predicting the color field gradient of a contact front, and finally predicting whether a contact front is going to break or not.  Using these predictions, our simulator recovers the overall shape of a liquid drop at every time step, and handles merging and splitting events by simple operations.  The experiment shows that our trained neural networks are able to perform predictions well.  The whole simulator is robust, convenient to use, and capable of generating realistic small-scale liquid effects in animation.
\end{abstract}



\maketitle
\thispagestyle{empty}
\section{Introduction}
\label{sec:intro}

Realistic simulation of small-scale liquid flows on solids, i.e., liquid drops and streamlets, is an important feature in virtual surgery, forensics (for simulating crime scenes), virtual painting, and entertainment applications.  Unfortunately, small-scale liquid flows on solids are notoriously difficult to simulate, because of the surface tension.  At a small scale, liquids, especially water, exhibit strong surface tension effect that places a stringent CFL condition on physics-based liquid simulation and demands a large computational cost to handle. While this issue can be remedied by ignoring interior flows and animating drop motions by the exterior surfaces only~\cite{Zhang:2012:DMS,Da:2016:SL}, a much more critical issue exists: how to model the surface tension among liquid, air, and solid. At the contact front where the three phases meet, the surface tension effect is highly complicated~\cite{Fowkes:1964:CAW,Mittal:2013:ACA}.  For example, it is known that the surface tension acts differently when the contact front is advancing vs. receding, known as the {\it hysteresis} effect. The surface tension also relies on the material properties of the solid, which are not conveniently measurable due to varying factors, including moisture, roughness, and dirt. In computer graphics, researchers~\cite{Wang:2005:WDS,Wang:2007:SGS} tried to simulate these liquid-solid effects by a varying contact angle scheme. While their result looks plausible, it is still far from realistic and they fail to display the same level of complexity demonstrated by real-world drops.

An interesting idea of simulating complex natural phenomena is to use data acquired from the real world. Real-world data are free of artifacts caused by incompetent physical models and they are able to capture complex dynamics well.  However, data-driven liquid simulation is not easy, due to the volatile nature of real-world liquids.
The recent advance in the machine learning technology inspires graphics researchers to revisit this idea from a learning-based perspective.
While the limited research in learning-based simulation is largely focused on generic simulation of large-scale liquid bodies, small-scale liquid flows and their interactions with solid surfaces remain unexplored.
Large-scale liquids are volatile and their flows are turbulent.  This causes many previous learning-based simulation techniques to be developed for predicting fluid behaviors in a local neighborhood, such as using convolutional neural networks (CNN)~\cite{Guo:2016:CNN,Yang:2016:DPM,Chu:2017:DSS}. In contrast, the motion of a small-scale liquid body, i.e., a liquid drop, is more predictable and often free of heavy topological events.  Hence we prefer to simulate small-scale liquids in a unique way, rather than in a unified way with large-scale liquids.

In this paper, we advocate the use of deep neural networks to predict the motion of a small-scale liquid body on surfaces as a whole.  We believe this is reasonable practice, because: 1) small-scale liquid flows are difficult to simulate in a physics-based way; 2) small-scale liquid flows are free of turbulence and they are more predictable.
We also would like to promote the idea of using a Lagrangian form to represent small-scale liquids. Compared with large-scale liquid bodies, small-scale liquids do not experience a significant number of merging or splitting events, therefore, we can afford handling them by simple operations.  The outcome of this research is a DNN-based simulator for animating small-scale liquids on solid surfaces. To this end, we have made the following technical contributions.
\begin{itemize}
\item{{\it Data preparation.} \hspace{0.16in} We present a convenient way to acquire raw liquid flow data from the real world.  We then develop practical techniques  to convert raw data into training data under a novel Lagrangian representation, which models an individual liquid drop solely by its contact front.}
\item{{\it Neural networks.} \hspace{0.16in}  Given the training data, we study the development and the training of three neural networks.  These networks are responsible for predicting the contour of a liquid drop's contact front, the color gradient at the contact front, and finally whether a contact front breaks at the next time step.}
\item{{\it DNN-based simulation.} \hspace{0.16in}  Our simulator combines neural networks with simple geometric/topological operations to achieve realistic simulation of water drops.  The implementation of this simulator involves a series of research on shape reconstruction, initialization, topological events, and flows on curved surfaces.}
\end{itemize}
Our experiment shows that the simulator is efficient, robust, parallelizable, convenient to use and its cost is linearly scalable to the number of liquid drops, as expected.  The animation results generated by our simulator are realistic and contain many complex details that cannot be easily handled by physics-based methods.

\section{Related Work}

\paragraph{Physics-based small-scale liquid simulation.} Physics-based liquid simulation has been extensively studied by graphics researchers in last two decades.  Here we narrow our discussions down to physics-based liquid simulation at a small scale.  Similar to general liquid simulation, the simulation of small-scale liquids can be developed in three ways: volumetric simulation~\cite{Enright:2002:HPL,Wang:2005:WDS,Wang:2007:SGS}, particle-based simulation~\cite{Akinci:2013:VST,He:2014:RSS,Chen:2015:WGP,Jones:2017:PDI}, and mesh-based simulation~\cite{Thurey:2010:MAM,Clausen:2013:SLS,Zhang:2012:DMS,Zhu:2015:CNF,Da:2016:SL}.  While researchers have expressed their strong interests in free liquid surface flows in these works, they paid much less attention to liquid flows on surfaces.  One of the reasons is because the surface tension at the contact front where liquid, air, and solid meet can be highly complicated.  In materials science and computational physics, this surface tension can be quantified by the contact angle between the liquid-air surface and the solid surface.  Wang and colleagues~\shortcite{Wang:2005:WDS} developed a virtual surface method to control the contact angles along the contact front in volumetric simulation.  This enabled them to model the hydrophobicity of a solid and produce contact angle hysteresis effects in animation. Later they~\shortcite{Wang:2007:SGS} extended this idea to the simulation of water drops on unstructured surface meshes. Recently, Zhang and collaborators~\shortcite{Zhang:2012:DMS} studied the virtual surface method in surface-only liquid simulation. In general, small-scale liquid flows on solids are still difficult to be simulated realistically and efficiently, not only because of the computational cost, but also because of the complex physical phenomena that cannot be precisely predicted by mathematical models.

\paragraph{Learning-based fluid simulation.}
The use of machine learning in fluid simulation starts to emerge recently, but it is still sparse in the computer animation community.
Ladicky and colleagues~\shortcite{Ladicky:2015:DFS} used random regression forests to correct particle positions in incompressible SPH simulation. Guo and collaborators~\shortcite{Guo:2016:CNN} trained a convolutional neural network (CNN) to efficiently approximate steady flow. Yang and colleagues~\shortcite{Yang:2016:DPM} accelerated a pressure projection solver in Eulerian fluid simulation by CNN. Tompson and collaborators~\shortcite{Tompson:2016:AEF} incorporated multi-frame information into a similar network structure for long-term accuracy improvement. Bonev and colleagues~\shortcite{Bonev:2017:PLS} proposed to predict surface deformation in water simulation by neural networks and soon Um and collaborators~\shortcite{Um:2017:LSM} used it for modeling water splashes.
To synthesize smoke animation, Chu and colleagues~\shortcite{Chu:2017:DSS} trained feature descriptors with CNN for fast example database query. As mentioned before, the difference between large-scale liquids and small-scale liquids restricts the applicability of these techniques in the simulation of small-scale liquid flows.

\section{Data Preparation}
Before we discuss the architecture of our deep neural networks in Section~\ref{sec:networks}, we would like to present the acquisition and the representation of our training data from real-world experiments.
The choices we make here are important to the success of the whole simulator and we will discuss the rationales behind them.

\subsection{Data Acquisition}
\label{sec:acquisition}
\begin{figure}
\centering
\includegraphics[width=3.35in]{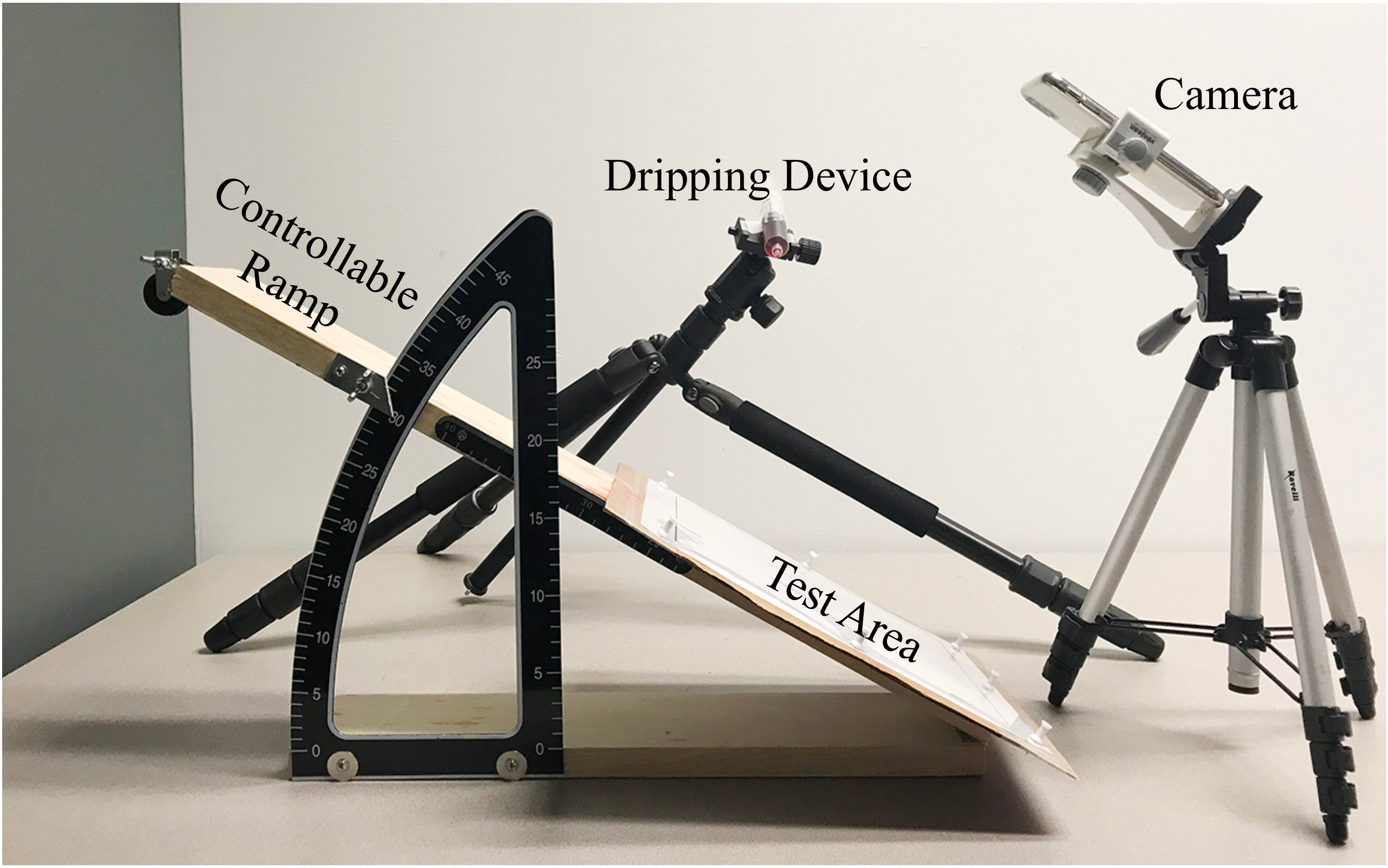}
\caption{Our data capture system. We build a simple system to capture liquid flow data from the real world.  This system is easy to build and convenient to use. }
\label{fig:datacapsetup}
\end{figure}

As mentioned in Section~\ref{sec:intro}, we choose to use real-world experiments, rather than physics-based simulation, to generate the training data, since they are more accessible and more accurate in describing liquid flow behaviors. The setup of our data capture system is shown in Fig. ~\ref{fig:datacapsetup}. It consists of a $8"\times12"$ test area made of a non-adsorbent matte plastic sheet, a dripping device that drops dyed water onto the top of the test area, and a commodity camera that records the video of liquid flows at 720p and 240 FPS. We set the test area on an inclined ramp with a fixed angle of $30^\circ$, so that liquid drops can flow at a suitable speed.
The surrounding environment uses indirect natural light to minimize reflected highlight spots on liquid drops.  We calibrate the camera and perform basic image rectification to remove camera distortion.

The length of each recorded video clip varies from 150 to 1200 frames.  In total, we capture 70 video clips, each of them contains one to three drops.
We typically do not include videos that contain too fast drops, since they can interfere with the overall network training.
Nearly one third of the clips contain splitting or merging events.  From the video clips, we obtain a data set that contains 80K training sets.  In addition to the training clips, we capture 10 more video clips for testing and evaluation purposes.

\subsection{Liquid Drop Representation}
Given the captured raw video data, we need to convert them into a compact representation for training purposes later in Section~\ref{sec:networks}.  Perhaps the most straightforward way is to cover the area of a liquid drop by a local window and use the interior color field as its representation.   However, such an Eulerian representation suffers from two drawbacks.  First, as the size of the liquid drop grows or shrinks, so does the size of the local window, which cannot be directly used for training neural networks.  Possible solutions include normalizing the window size, or using a fixed window size.  But they can cause other issues, such as wasted window space or confusing networks by liquid drops in different sizes.  The second drawback is the dimensions of this representation. For a small liquid drop contained in a low-resolution window with $32 \times 32$ cells only, the overall dimension of the Eulerian representation vector would be 1024.  This is too big for neural networks to be trained with efficiency, especially if the training set is large.

Instead, we present a Lagrangian liquid drop representation, solely based on the drop's contact front as shown in Fig.~\ref{fig:video2contour}. This representation uses a fixed number of control points at the contact front, each of which stores its location and its color gradient in 2D.  The point locations outline the contour of the liquid drop, and the sampled gradients determine the interior shape of the drop.
Our representation is not only compact, but also effective in separating the 2D contour shape from the 3D interior shape. This allows the contour and the interior, i.e., the locations and the gradients, to be trained and predicted by two separate networks, as shown in Section~\ref{sec:networks}.

\subsubsection{Contour extraction}
\label{sec:contour_ext}

Given a video sequence obtained by our data capture device, our first job is to extract the contours of the contact fronts for our learning system.

\begin{figure}[t]
	\centering
	\subfigure[Image]{\includegraphics[width=0.79in]{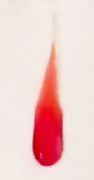}\label{fig:lorlem1}}\hfill
	\subfigure[Binary image]{\includegraphics[width=0.79in]{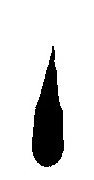}\label{fig:lorlem2}}\hfill
	\subfigure[Dense samples]{\includegraphics[width=0.79in]{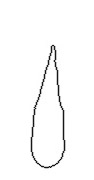}\label{fig:lorlem3}}\hfill
	\subfigure[Control points]{\includegraphics[width=0.79in]{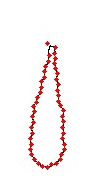}\label{fig:lorlem4}}\hfill
	\vspace{-0.08in}
	\caption {The contour of a liquid drop in a Lagrangian representation. After we densely sample the contour of a liquid drop in a binary image (b), we approximate it by a B-spline curve with a fixed number of control points in (d). These control points form a vector representing the liquid drop contour, used by our training system. }
	\label{fig:video2contour}
\end{figure}

First, we apply morphology operations to smooth out any irregularity or noise in each frame, and run Otsu's thresholding algorithm to convert the frame into a binary image, as shown in Fig.~\ref{fig:video2contour}b. We then use an active contour method on the binary image to obtain a contour, represented as a densely sampled piecewise linear curve shown in Fig.~\ref{fig:video2contour}c.  To make those contours useful in our training system, we need both their shapes and their temporal dependencies. While a contour in one frame remains as a contour in the next frame most of the time, merging and splitting events do occur occasionally as well.
Our solution is an automatic contour tracking algorithm that uses contour overlaps to determines whether a contour in one frame becomes:
\begin{itemize}
\vspace{-0.02in}
\item{another contour in the next frame,}
\vspace{-0.02in}
\item{or part of another contour in the next frame, due to merging,}
\vspace{-0.02in}
\item{or two contours in the next frame, due to splitting.}
\vspace{-0.02in}
\end{itemize}
We do not consider the splitting of a contour into three or more contours, nor the merging of three or more contours, since they are rare given the frame rate of our video camera.
The outcome of our algorithm is a set of $N$ contour sequences, each of which is tracked over a different number of frames in our videos.  The contour sequences starts when the video starts, or when it gets newly generated by merging or splitting event. The contour sequence ends when the video ends, when it leaves the field view, or when it gets terminated by merging or splitting events. Those contour sequences that end with splitting are specifically labeled for breakage prediction later in Subsection~\ref{sec:breakage}.

The dense samples can represent the contour well, but they are not suitable for training neural networks, since the number of samples per contour is determined by the contour size, which can vary from contour to contour, and from time to time. To solve this issue, we use a cubic B-spline curve with exactly 52 control points to approximate the sampled piecewise linear contour. The control points are ordered in a clock-wise fashion from the top.  In this way, we represent each contour by a control point vector in the $\mathbb R^{104}$ space.  Fig.~\ref{fig:video2contour}e shows that the reconstruction from the control points is a good approximation to the original contour.

\subsubsection{Gradient extraction}
\begin{wrapfigure}{r}{1.2in}
\vspace*{-.14in}
\includegraphics[width=1.2in]{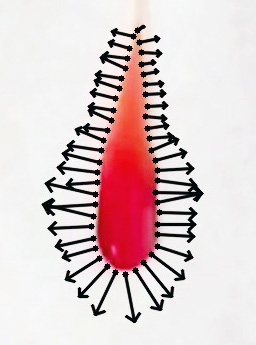}
\vspace*{-.18in}
\caption{The gradients at the control points of a contact front.}
\vspace{-.16in}
\label{fig:grad}
\end{wrapfigure}
The contour representation given in Subsection~\ref{sec:contour_ext} does not provide any information inside of the contour, i.e. the color intensity of every liquid pixel.
Based on the assumption that the surface tension is sufficiently large to smooth out the mean curvature of the liquid surface, we assume that the color field is smooth as well and it can be  estimated from its gradient at the contact front as discussed later in Section~\ref{sec:recover}.

To predict such gradient information by our neural networks, we extract the gradient of the color field at each control point by the Sobel operator in the image space. We stack the gradient vectors together to form another large vector for training gradient prediction networks.
Fig. ~\ref{fig:grad} shows the gradients extracted at the contact front of a liquid drop. Compared with the contour, the gradients are more sensitive to image noises and imperfections, such as highlight spots.  While these issues can be addressed by applying more image editing operations, we find it is unnecessary as we expect the neural networks to learn the noises and remove their influence on predictions automatically.

\section{Neural Networks}
\label{sec:networks}

Given the extracted contour and gradient at the contact front of a liquid drop, our neural networks are responsible for predicting the contour and the gradient at the next time step.
It also needs to predict whether the contact front breaks or not, for simulating liquid drop splitting events in Subsection~\ref{sec:merging}.

\begin{table*}
\subfigure[Contour prediction subnet]
{
\begin{tabular}{cc|c }
\hline
 \rowcolor{gray!50}
X Coordinates & Y Coordinates & Center \\
\hhline{---}
Input--52 & Input--52 & Input--2 \\
LSTM(Lin)--260 & LSTM(Lin)--260 & LSTM(Lin)--260 \\
\hhline{--}
\multicolumn{2}{ c| }{ \cellcolor{gray!50} Merge} & LSTM(Lin)--260\\
\cline{1-2}
\multicolumn{2}{ c| }{LSTM(Lin)--260} & LSTM(Lin)--260 \\
\multicolumn{2}{ c| }{LSTM(Lin)--260} & \\
\multicolumn{2}{ c| }{LSTM(Lin)--260} & \\
\multicolumn{2}{ c| }{LSTM(Lin)--260} & \\
\hline
 \rowcolor{gray!50}
\multicolumn{3}{ c }{Merge} \\
\hline
\multicolumn{3}{ c }{LSTM(Lin)--260} \\
\multicolumn{3}{ c }{LSTM(Lin)--260} \\
\hline
\multicolumn{3}{ c }{Dense(Lin)--106} \\
\hline
\vspace{-0.06in} \\
\end{tabular}
}
\hspace{0.2in}
\subfigure[Gradient prediction subnet]
{
\raisebox{-0.24in}{
\begin{tabular}{ c }
\hline
 \rowcolor{gray!50}
Gradient Magnitude \\
\hline
Input--104 \\
LSTM(Lin)--250  \\
LSTM(Lin)--250  \\
LSTM(Lin)--250  \\
LSTM(Lin)--250  \\
LSTM(Lin)--250  \\
LSTM(Lin)--250  \\
\hline
Dense(Lin)--50 \\
\hline
\vspace{-0.06in} \\
\end{tabular}
}}
\hspace{0.2in}
\subfigure[Breakage prediction subnet]
{
\raisebox{-0.24in}
{
\begin{tabular}{ c}
\hline
 \rowcolor{gray!50}
X and Y Coordinates \\
\hline
Input--50 \\
Dense(ReLu)--150  \\
Dense(ReLu)--150 \\
Dense(ReLu)--150 \\
Dense(ReLu)--150 \\
Dense(ReLu)--150 \\
Dense(ReLu)--150 \\
\hline
Dense(Sig)--1 \\
\hline
\vspace{-0.06in} \\
\end{tabular}}
}
\vspace{0.06in}
\caption{The structure of three subnets. We use the following name convention for each layer: {\it type (\it activation)}--{\it dimension}. For instance, {LSTM(Lin)}--260 stands for a LSTM cell layer with 260 dimensions and linear activation, and {Dense(Lin)}--104 represents a fully connected dense input layer with 104 dimensions and linear activation.  All of the LSTM layers except for the last one generates a sequence output.}
\label{tab:struct}
\end{table*}

Our networks are a combination of fully connected classification networks and recurrent regression networks.
Our deep learning architecture is a chimeric network consisting of three individual sub networks (hence abbreviated as \textit{subnets}) operating at each individual time step.
To reduce the overfitting issue, we adopt the dropout approach in all of the layers except for the output layer. What it basically does is to randomly select a network node and exclude it during the training stage.  In this way, the networks can be more generalized without developing co-dependencies among its neurons. See~\cite{Srivastava:2014:DSW:2627435.2670313}  for more details.

\subsection{Contour Prediction Subnet}
\label{sec:contour}
Given the contours of a contact front at the previous $K$ time steps, we would like to predict the contour at the next time step by a subset, rather than physics models.
The training process of such a subnet involves heavy temporal data dependencies, which are prone to problems of exploding and vanishing gradients.
To avoid these problems, we choose to use a recurrent neural network (RNN) architecture. Based on the compact representation provided in Subsection~\ref{sec:contour_ext}, we can then consider the subnet training process as predicting the temporal trajectory of contour control points.

Our deep learning model uses long short term memory (LSTM) units~\cite{Hochreiter:1997:LSM}, based on the recurrent neural network (RNN) architecture. LSTM is popularly used in natural language processing community to model complex language model dependencies. As is the common practice for most deep learning applications, we normalize (with respect to the overall dimensions of the 2D domain on which the B-spline curve is drawn) and centralize the control points. As an additional input, we track and predict the center using the previous contours.  Considering the complexity and the variability of our network, its structure is shown in Table ~\ref{tab:struct}a. We note that each input has its own separate processing as well as combined processing.
This allows all of the attributes of this nonlinear input function to be learned both on its own and with mutual dependencies.  To optimize network parameters, we use stochastic gradient descent optimization, with Nesterov acceleration~\shortcite{Nesterov:AMU} to speed up the convergence rate.  We have also tested the adaptive moment estimation (Adam) technique~\cite{Kingma:2014:ADA}, but its performance is generally lower than stochastic gradient descent with Nesterov acceleration. The convergence of the training process for this contour prediction subnet is illustrated in Fig.~\ref{fig:plot}a.

\subsection{Gradient Prediction Subnet}
\label{sec:gradient}
\begin{figure*}
\centering
\subfigure[Contour prediction subnet]{\includegraphics[width=2.32in]{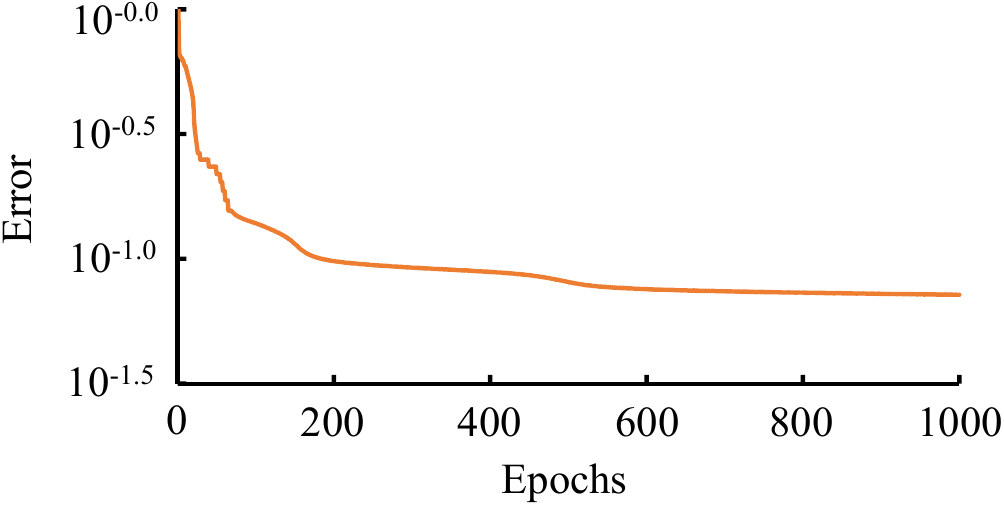}}
\subfigure[Gradient prediction subnet]{\includegraphics[width=2.32in]{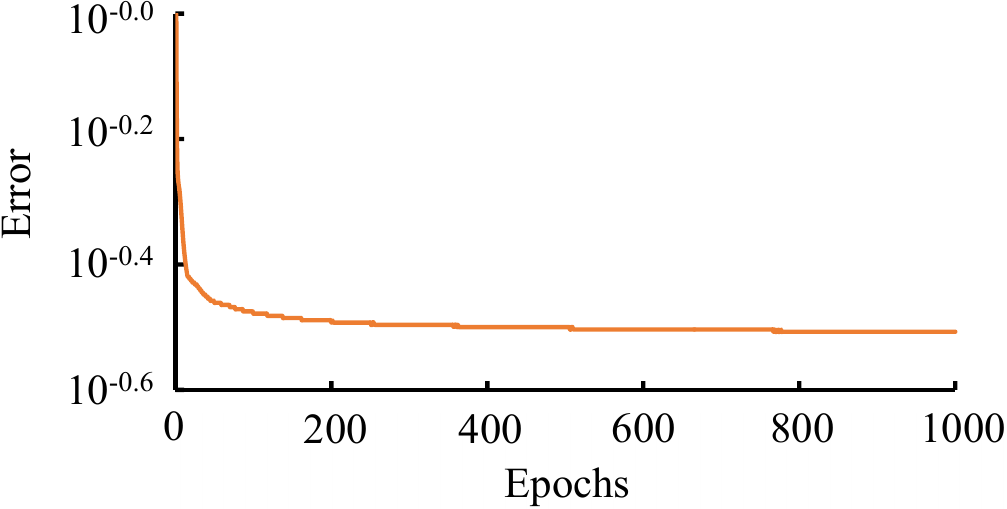}}
\subfigure[Breakage prediction subnet]{\includegraphics[width=2.32in]{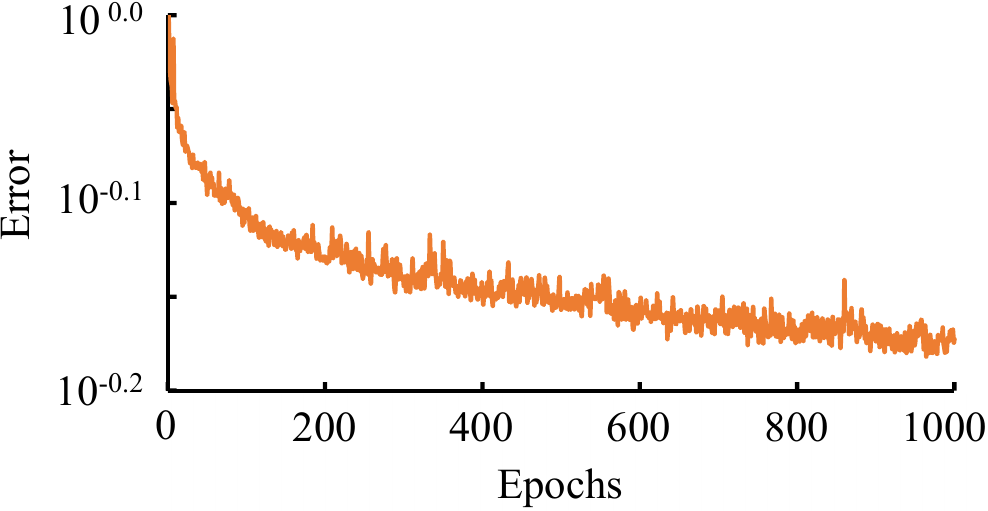}}
\vspace{-0.1in}
\caption{Relative training error losses during the training of the three contour prediction subnets.
For contour and gradient predictions, the error is defined as the L2-norm between the ground truth and the prediction outcome.
For breakage predictions, the error is defined as the accuracy of the prediction outcome.}
\label{fig:plot}
\end{figure*}

To fill the interior of the contact front, we need a subnet to predict the gradient of the color field at the contact front next.
Similar to the contour prediction subnet in Subsection~\ref{sec:contour}, this gradient prediction subnet uses the gradients at the last $K$ time steps and it is formulated under a LSTM-based RNN framework.  To simplify the training process, we predict the gradient magnitude at each control point only and we assume that the gradient is always perpendicular to the B-spline curve.  The structure of the gradient prediction subnet is shown in Table ~\ref{tab:struct}c.  The training procedure is similar to that of the shape prediction subnet. The convergence behavior of the training process for the contour prediction subnet is illustrated in Fig.~\ref{fig:plot}b.

\subsection{Breakage Prediction Subnet}
\label{sec:breakage}
Finally, we use a breakage prediction subnet to determine whether the contact front of a liquid drop breaks at the next time step or not.  If it does break, it will cause the liquid drop to split into two new drops.  While this phenomenon has a substantially complex explanation in the real world, we formulate it in our simulator as a binary classification problem based on the contact front information.

The input to this breakage prediction subnet is the mean-centered and normalized control points and its decision is binary: whether the contact front breaks or not. For this classification problem, we use a fully connected deep learning classifier network. The structure of the breakage prediction subnet is shown in Table ~\ref{tab:struct}c.

An important aspect to note here is that breakage is a relatively uncommon event and the resulting class classification problem is highly unbalanced.  Therefore, accuracy is not a good measure of the overall classification success.  Instead, we train the network by a balanced mixture of positives (those with breakages) and negatives (those without breakages), which are selected using an under-sampling method based on the near-miss algorithm~\cite{Zhang03}.

Unlike the regression problems discussed in Subsection~\ref{sec:contour} and~\ref{sec:gradient}, this classification problem can use either adaptive moment estimation (Adam)~\cite{Kingma:2014:ADA} or stochastic gradient descent with Nesterov acceleration, both of which have plausible convergence rates as shown in our experiment.  In our simulator, we choose Adam and the convergence of the training process is shown in Fig.~\ref{fig:plot}c.

\section{DNN-based Simulation}
\label{sec:simulation}
Given the neural networks trained for predicting the contour, the gradient, and the breakage of a contact front, we now would like to develop a simulator that animates the time evolution of every liquid drop.  To begin with, we will discuss the 3D shape reconstruction of a liquid drop from its contact front in Subsection~\ref{sec:recover}. We will then study how to initialize the simulation of a liquid drop newly added to the scene in Subsection~\ref{sec:initialize}. In Subsection~\ref{sec:merging}, we investigate the modeling and simulation of topological events, i.e., splitting and merging of liquid drops. Finally, we will present an ad hoc method for simulating drops on curved surfaces in Subsection~\ref{sec:curve}.

\subsection{Shape Reconstruction}
\label{sec:recover}

\begin{figure}
\centering
\includegraphics[width=3.3in]{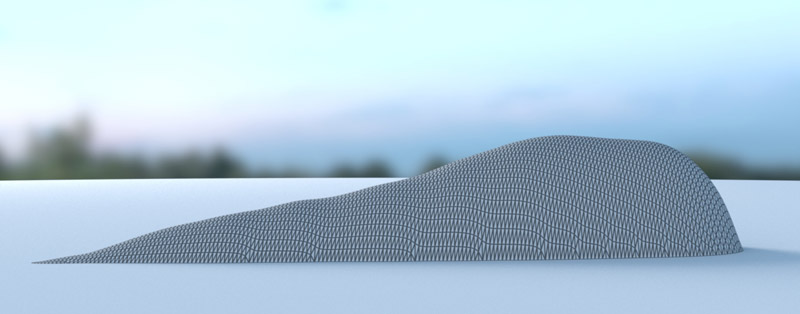}
\caption{The side view of a liquid drop. Using the gradient information at the contact front, we solve a biharmonic equation to reconstruct the overall shape of this liquid drop.}
\label{fig:side}
\end{figure}

Without external force, a liquid drop in its equilibrium state has a uniform mean curvature over its surface. With external force, a moving liquid drop no longer has a uniform mean curvature surface, but the surface should still be sufficiently smooth.  Similar to~\cite{Wang:2007:SGS}, we describe such a smooth liquid surface by a biharmonic equation of the color field, subject to boundary conditions at the contact front:
\begin{equation}
\left\{
\begin{array}{cl}
\nabla^4 c(\mathbf x) = 0,  & \mathrm{for\;} \mathbf x \in \Omega,  \\
 c(\mathbf x) = 0, & \mathrm{for\;} \mathbf x = \partial \Omega, \\
\nabla c(\mathbf x) = \mathbf g(\mathbf x),  & \mathrm{for\;}  \mathbf x \in \partial \Omega,
\end{array} \right.
\label{eq:biharmonic}
\end{equation}
in which $\Omega$ is the liquid drop domain encircled by the contact front and $\mathbf g(\mathbf x)$ is the predicted gradient at the contact front.  To solve the PDE system in Equation~\ref{eq:biharmonic}, we discretize the biharmonic operator by finite differencing and creates a stencil covering the 2-ring neighborhood of each grid cell.  To make this stencil well defined near the boundary, we use the two boundary conditions to generate the color values around the contact front.  The boundary conditions are discretize in the first order.
The resulting linear system is denser than the linear system of Laplace's equation, but still sparse enough for fast linear solvers.  Therefore, we solve it efficiently on the CPU by MKL PARIDSO in real time. Once we obtain the color field, we convert it into a height field after proper scaling for maintaining an explicitly tracked volume, and generate a triangle mesh from the height field for visualization afterwards.

Fig.~\ref{fig:side} shows the side view of a liquid drop generated by our method.  Thanks to larger gradients near the head and smaller gradients near the tail, we are able to recover a plausible shape of this liquid drop. We note that first-order boundary conditions can cause grid artifacts at both the contact front and on the interior surface. Instead of switching to use higher-order boundary conditions, we simply smooth the contact front and the interior surface as a post-processing.

\subsection{Initialization}
\label{sec:initialize}

To initialize the simulation of a liquid drop, we request the contour of its contact front to be provided externally, either by animators or from a contour template database.  One problem is that the contour prediction network needs the contours at the last $K$ time steps, which do not exist at the very beginning of the simulation process.  To solve this problem, we simply perform a {\it cold start}, by replicating the same contour shape $K-1$ times.  This is a common practice in LSTM, when it is applied to language models and recommender systems.


Similar to shape prediction, gradient prediction also needs the last $K$ color field gradients to begin with.  Simply asking the animators  to provide such gradient information is impractical.  One solution is to initialize the whole 3D shape of a liquid drop and then calculate the contour and the gradient at the contact front from the shape.  Alternatively, we provide a more convenient solution, which automatically chooses the proper value of each gradient vector from captured real-world data.  Our idea is based on the assumption that liquid drops with similar contact front contours have similar contact front gradients.  We search through a representative liquid drop database and find the one with the most similar contact front contour.  We then assign its gradients to the newly initialized liquid drop.  We note that this database is used for initialization only, so it is much more compact than the training database.


\subsection{Splitting and Merging}
\label{sec:merging}
In this subsection, we will discuss how to process two topological events in our simulation: splitting and merging.

\subsubsection{Splitting}
\label{sec:splitting}
Once we are informed by the breakage prediction subnet about a splitting event, the first thing we must decide is: where does splitting happen? Assuming that splitting always happens at the ``neck" of a liquid drop, we propose to find the splitting location by solving a discrete optimization problem:
\begin{equation}
\left\{i, j \right\} = \arg\min \left\| \mathbf x_i - \mathbf x_j \right\| - C_{i,j}, \quad \mathrm{for}\; \mathbf n_i \cdot \mathbf n_j <\delta.
\label{eq:opt}
\end{equation}
in which $\mathbf x_i$ and $\mathbf x_j$ are the 2D positions of two control points $i$ and $j$, $C_{i,j}$ are their curvalinear distance along the contact front, and $\mathbf n_i$ and $\mathbf n_j$ are their normals.
Intuitively, the first term in Equation~\ref{eq:opt} tries to reduce the Euclidean distance between the two points, while the second term tries to make them well separately along the contact front, and the constraint ensures that the two points are well opposite to each other.  Since there are only 52 control points per contour, we can check every control point pair to find the one with the minimum cost, with some culling to avoid unnecessary computation.

Once we decide the splitting pair, we simply divide the B-spline curve into two and resample each new curve by their own control points. We then calculate the gradients at these new control points by linearly interpolating the gradients of the original control points. We note that this resampling and interpolation process happens not only at the current time step, but also at the last $K-1$ time steps.  This ensures that the newly generated drops are ready for future prediction.

\subsubsection{Merging}
Compared with splitting, merging is relatively easy to handle in our simulator.  At every time step, we first check whether the B-spline contours of two liquid drops start to overlap using dense samples. If they do, we remove the intersecting samples and approximate the remaining samples by a new B-spline curve. The gradients at newly generated control points are linear interpolated from the old ones, as in Subsection~\ref{sec:splitting}.

An interesting issue is how to initialize the needed data of this new drop at the previous $K-1$ time steps.  It makes little sense to simply combine the information of the two merged drops, since they are irrelevant before merging happens and the prediction result would be problematic as shown in our experiment.
Instead we simply treat a merged drop as a new drop and initialize it by a "cold start". A side effect of this practice is a small lag right after merging.  One possible solution is to calculate the average control point displacements at the last $K-1$ time steps and apply them to the whole contour of the new drop to obtain the estimations of the last $K-1$ contours. We have not tried this idea yet.

\subsection{Liquid Flows on Curved Surfaces}
\label{sec:curve}
Our data capture device acquires the original data as liquid drops flowing on a planar slope with a fixed incline angle.  While the resulting neural networks can effectively predict liquid behaviors in the same scenario, they become less accurate in the simulation of liquid drops on curved surfaces, or planar slopes with other incline angles.
\begin{figure}
\centering
\subfigure[The whole scene]{\includegraphics[width=3.32in]{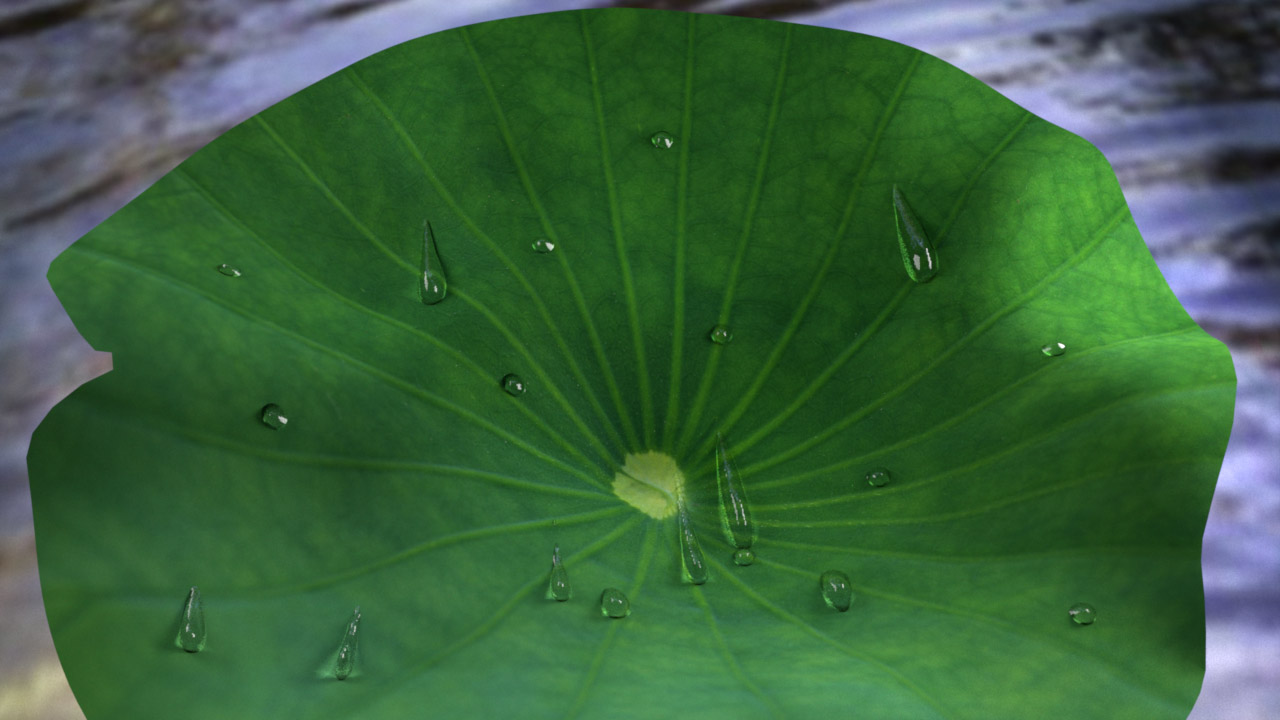}}
\subfigure[A static drop]{\includegraphics[height=1.69in]{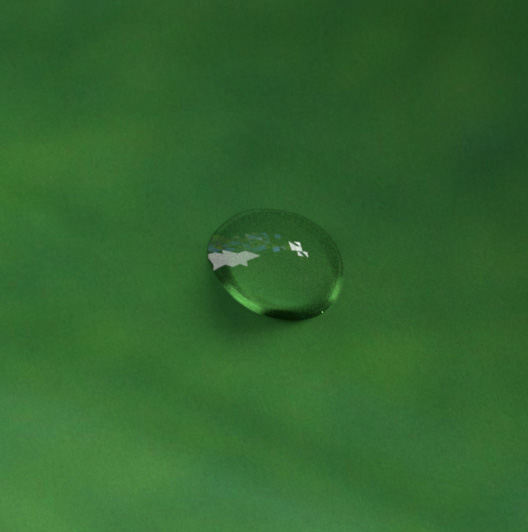}}
\subfigure[A dynamic drop]{\includegraphics[height=1.69in]{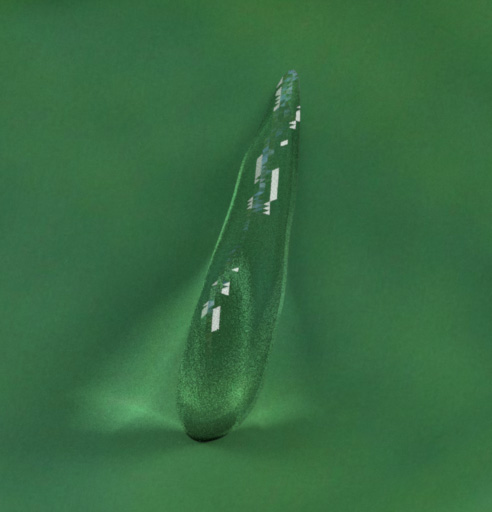}}
\vspace{-0.1in}
\caption{A lotus leaf example.  Our simulator can simulate both static and dynamic liquid drops on a curved solid surface.}
\label{fig:leaf}
\end{figure}

To solve this problem, we propose to provide two adjustments to our simulator.  First, we discretize the curved surface into a height field as well and place the height field of the liquid drop on it.  Doing this allows the generated liquid triangle mesh to be aligned with the curved surface.  Second, we scale the contact front and gradient of a liquid drop by a factor: $s = (sin \theta)^{1/3}$ during prediction, in which $\theta$ is the average incline angle of the surface area contacting the liquid drop.  The motive behind this practice is that larger drops move faster while smaller drops move slower, as shown in Fig.~\ref{fig:four}.  By adjusting the size of the liquid drop during prediction, we can indirectly control its moving speed due to a varying incline angle.  Fig.~\ref{fig:leaf} shows that these techniques can be used to effectively simulate liquid flows on a curved lotus leaf.

\section{Results}
(Please watch the supplemental video for animation examples. We will release our 10G data set including recorded videos and pre-trained models after the paper gets published.)
Our framework is implemented in Python 3 using Keras with the TensorFlow backend. Each network is trained for 1000 epochs with a mini-batch size of 128. The initial learning rate is $10^{-2}$ and learning rate decay of $10^{-6}$ was used. All the training procedures are done on an Intel Xeon E6-2640 Processor with a single NVIDIA P100 GPU. The evaluation is done on a desktop with Intel i7 Processor with NVIDIA  980GTX GPU. It takes approximately 0.5s per epoch for the breakage prediction subnet, 204s per epoch for the shape predictor subnet and 130s per epoch for the gradient predictor subnet to train respectively. The total training time is approximately four days.

\begin{figure}
\centering
\subfigure[Before simulation]{\includegraphics[width=1.65in]{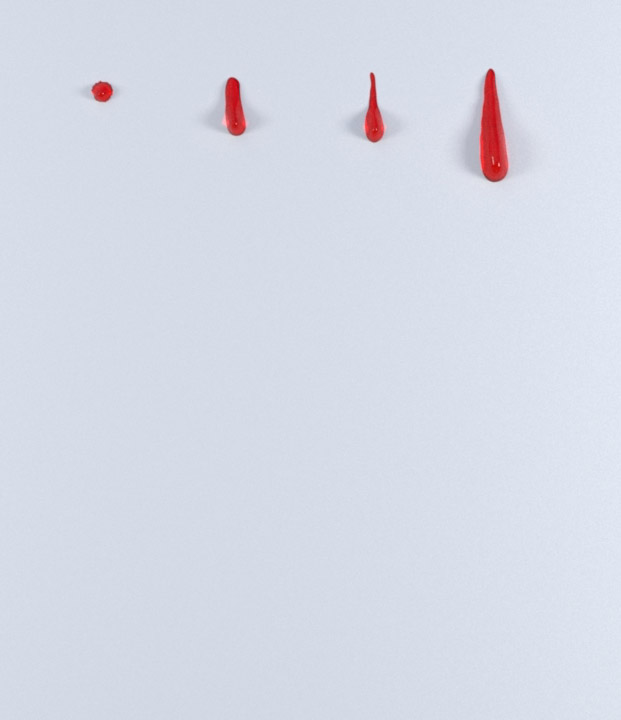}}
\subfigure[After simulation]{\includegraphics[width=1.65in]{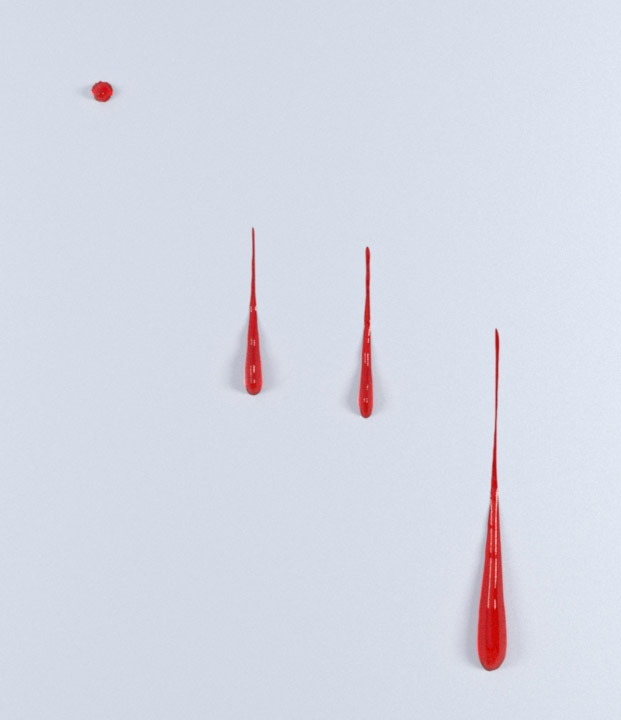}}
\vspace{-0.1in}
\caption{Four liquid drops in different sizes. This example demonstrates the ability of our simulator in animating flowing behaviors of liquid drops, based on their sizes.}
\label{fig:four}
\end{figure}

\begin{figure}
\centering
\subfigure[Ground truth]{\includegraphics[width=0.81in]{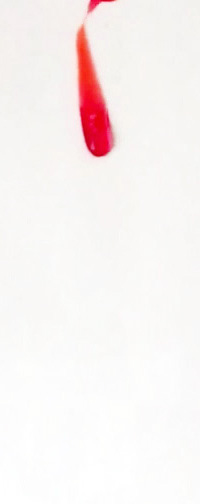}}
\subfigure[Simulated result]{\includegraphics[width=0.81in]{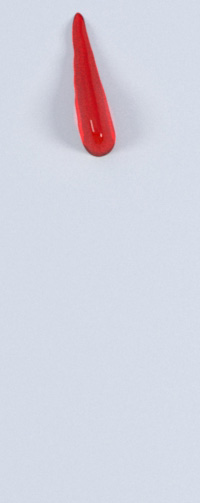}}
\subfigure[Ground truth]{\includegraphics[width=0.81in]{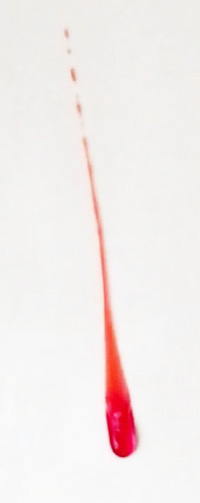}}
\subfigure[Simulated result]{\includegraphics[width=0.81in]{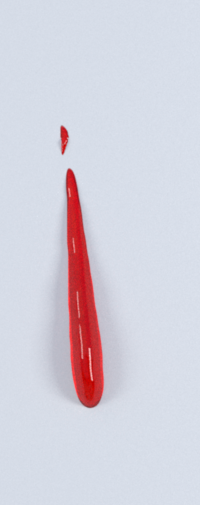}}
\vspace{-0.1in}
\caption{A comparison example with the ground truth.  The ground truth here is a video clip selected from the evaluation data set, which is unused for training.}
\label{fig:groundtruth}
\end{figure}

\paragraph{Flow comparison with the ground truth.}\hspace{0.16in}  To evaluate how accurate our simulation is, we select one liquid drop from the evaluation data set and use our simulator to generate its time evolution from the initial state, as shown in Fig.~\ref{fig:groundtruth}.  Fig.~\ref{fig:groundtruth} shows that our simulator can provide a plausible prediction for the dynamic motion of this drop.  The breakage prediction subset also correctly predicts one splitting event, although the real liquid drop splits multiple times.  We think there are several factors contributing to the difference between our simulation and the ground truth video.  First, we use the color image to estimate the initial drop state, which is subject to noises and errors. Second, our neural networks have their own errors.  Third, the shape reconstruction process that generates the 3D shape of the liquid drop may not be accurate.  Finally, each video is captured at slightly different physical conditions, which can cause the same drop to flow differently anyway.

\paragraph{Flow comparison of drops in different sizes.}\hspace{0.16in}   Fig.~\ref{fig:four} compares four drops in different sizes flowing on the same solid surface.  It shows that our simulator can correctly handle flowing speeds with respect to drop sizes. In particular, large drops flow faster, while small drops remain static because of contact angle hysteresis. When two drops are in similar sizes, such as the two in the middle of Fig.~\ref{fig:four}a, it becomes difficult to know which one moves faster ahead of time and our simulator determines the outcome in an uncontrollable fashion.

\subsection{Limitations}
\label{sec:limitations}
Due to a limited number of data categories, our neural networks are not able to predict flow behaviors for a wide range of liquid and solid material properties.  Even if they do, our intrinsic data representation is unable to handle highly hydrophobic surfaces with contact angles greater than 90$^\circ$.  Currently, our simulator provides only an approximation to liquid flows on curved surfaces and the result would be problematic if the surface is highly uneven.  The simulator has difficulty in handling liquid flows under external forces, such as wind or user interaction. The simulator becomes less accurate when it simulates long liquid streamlets and their topological events.  Since the simulator is ignorant of volume preservation, it needs the volume to be explicitly tracked and maintained.  Finally, our current simulator works as it is and it cannot handle liquids entering or leaving solid surfaces, such as the dripping effect.

\section{Conclusions and Future Work}
In this paper, we present a novel approach to simulate small-scale liquid flows on solid surfaces by deep neural networks.  Our research shows that the shape and the topological events associated with an individual liquid drop can be well predicted by trained neural networks. Combined with geometric operations, these networks can be effectively used to form a learning-based simulator for realistic animation of small-scale liquids.

Our future plan is centered around solving the limitations summarized in Subsection~\ref{sec:limitations}.
Specifically, we plan to collect more diversified data, to see if our simulator is able to simulate flow behaviors for various liquid and solid material properties.  We also plan to combine our learning-based system with existing physics-based liquid simulation techniques for simulating comprehensive liquid behaviors.  How to parallelize our system for real-time liquid simulation, especially on the GPU, is another important problem we would like to investigate.

\begin{acks}
This work was funded by NSF grant CHS-1524992. The authors would also like to thank Adobe
Research and NVIDIA Research for additional equipment and funding supports.
\end{acks}

\bibliographystyle{ACM-Reference-Format}
\bibliography{neuraldrop}
\end{document}